\documentclass[11pt,a4paper]{article}
\usepackage{jheppub_kim}

\usepackage{pdflscape}
\usepackage{amsmath}
\usepackage{amssymb}
\usepackage{dcolumn}
\usepackage{bm}
\usepackage{color}
\usepackage{epsfig}
\usepackage{amsfonts}
\usepackage{graphicx}
\usepackage{subfigure}
\usepackage{dcolumn}

\begin{document}

\title{Quantum Corrections to a Finite Temperature BIon}
\author[a]{Behnam Pourhassan,}
\author[b]{Sanjib Dey,}
\author[c]{Sumeet Chougule,}
\author[d]{Mir Faizal}

\affiliation[a]{Iran Science Elites Federation, Tehran, Iran}
\affiliation[a]{Canadian Quantum Research Center 204-3002 32 Ave Vernon, BC V1T 2L7 Canada}
\affiliation[b]{Department of Physical Sciences, Indian Institute of Science Education and Research Mohali, Sector 81, SAS Nagar, Manauli 140306, India.}
\affiliation[c]{Centro de Estudios Cient\'{\i}ficos (CECs), Av. Arturo Prat 514, Valdivia, Chile}
\affiliation[c]{Departamento de F\'{\i}sica, Universidad de Concepci$\acute{o}$n, Casilla 160-C, Concepci$\acute{o}$n, Chile}
\affiliation[d]{Irving K. Barber School of Arts and Sciences, University of British Columbia, Kelowna, British Columbia, V1V 1V7, Canada.}
\affiliation[d]{Department of Physics and Astronomy, University of Lethbridge, Lethbridge, Alberta, T1K 3M4, Canada.}

\emailAdd{b.pourhassan@candqrc.ca}
\emailAdd{dey@iisermohali.ac.in}
 \emailAdd{chougule@cecs.cl}
 \emailAdd{mirfaizalmir@googlemail.com}

\abstract{In this paper, we will analyze a  finite temperature BIon, which is a finite temperature brane-anti-brane wormhole configuration.
We will analyze the quantum fluctuations to this BIon solution using the Euclidean quantum gravity.
It will be observed that these quantum fluctuations produce logarithmic corrections to the entropy of this finite temperature BIon solution.
These corrections to the entropy also correct the internal energy and the specific heat for this finite temperature BIon.
We will also analyze the critical points for this finite temperature BIonic system, and analyze the effects of quantum corrections
on the stability of this system.}


\maketitle

\section{Introduction}
In string theory, it is possible to analyze certain physical objects
in a region of spacetime in terms of very different objects.
Thus, it is possible to analyze a system of many coincident strings
in terms of D-brane geometry, and this is done in the BIon solution
\cite{Callan_Maldacena, Gibbons}.
So, this  BIon solution can describe an   F-string
coming out of the D3-brane or  a D3-brane  parallel to an
anti-D3-brane, such that they are connected by a wormhole
with F-string charge. This configuration is called as the
 brane-antibrane-wormhole configuration.
It is also possible to use such a solution to analyze D-branes
probing a thermal background \cite{BIon01, bion}.
This can be done using the blackfold approach \cite{q1, q2, q5, q4}.
In this method,  a large number
of coincident
D-branes form a  brane probe. Furthermore, as
this  probe is in thermal equilibrium with the background, this method
has been used to heat up a BIon. This was done by putting
it in a hot background. It is also possible to analyze the thermodynamics
of this finite temperature BIon solution   \cite{BIon01, bion}.
In this paper, we will analyze the effects of thermal fluctuations
on the thermodynamics of this system.

The   entropy-area law   of black holes thermodynamics
\cite{Hawking,Bekenstein}, is expected to get modified near Planck scale due to quantum fluctuations \cite{Rama}.
These quantum fluctuations in the geometry of  any
black object are expected to produce thermal fluctuations in its associated
thermodynamics.
It is interesting to note that the thermal fluctuations produce a logarithmic correction term to the thermodynamics of
 black objects \cite{1, 2, 3, 3a}.  The consequences of such logarithmic correction have been studied for
a  charged AdS
black hole \cite{12}, charged hairy black hole \cite{12a}, a black saturn \cite{13}, a Hayward black hole \cite{18}   a small singly spinning Kerr-AdS black hole \cite{19}, and
a dyonic charged AdS black hole \cite{4}.
In non-perturbative quantum general relativity, the density
of microstates were associated with the conformal blocks  has been used to obtain  logarithmic corrections to the entropy  \cite{Ashtekar}.
It has also been  demonstrated that  the Cardy formula can produce logarithmic correction terms
for all black objects whose microscopic degrees of freedom are characterized by a conformal field theory \cite{Govindarajan}. The logarithmic correction
has also been studied  from the black hole in the presence of matter fields \cite{Mann_Solodukhin} and dilatonic black holes \cite{Jing_Yan}.
Leading order quantum corrections to the semi-classical black hole entropy have been obtained \cite{5},
and applied to   G\"{o}del black hole \cite{6, koka2}.
The logarithmic corrections were also used to study different  aspects of  regular black holes satisfying the weak energy condition \cite{10},
three-dimensional black holes with
soft hairy boundary conditions \cite{new2}, and certain aspects of Kerr/CFT correspondence \cite{11}.
The logarithmic corrected entropy also corrects some hydrodynamical quantities, and so the
the field theory dual to such corrected solutions has also been studied  \cite{23,24,27,28,29}.

The logarithmic corrections to the entropy of various black holes have also been obtained
using the Euclidean Quantum Gravity \cite{32, 8, 9}. In this approach, Euclidean Quantum Gravity \cite{hawk}  is used to
obtain the  partition function  for the black hole, which is then  used to obtain the logarithmic corrections to the
thermodynamics of that black hole.
As the logarithmic corrections occur almost universally in the thermodynamics of black objects,
in this paper we will analyze the consequences of such corrections for a thermal BIon.
 We compute the quantum  correction to the black hole entropy, internal energy,
specific heat using the Euclidean Quantum Gravity  \cite{hawk}.
We find that the logarithmic correction affects the critical points, and the corrections  significantly  change the stability of this system.
\section{Euclidean Quantum Gravity}
Now, we start with  the Euclidean Quantum Gravity that
is obtained by performing a Wick rotation on the temporal coordinates in the   path integral. Thus, we obtain
   gravitational partition function in Euclidean Quantum Gravity \cite{hawk},
\begin{equation}\label{PF}
Z = \int [\mathcal{D}]e^{-\mathcal{I}_\text{E}}= \int_0^\infty \rho(E)e^{-\beta E} \text{d}E,
\end{equation}
where $\mathcal{I}_\text{E}$ is the Euclidean action for the BIon solution   \cite{BIon01, bion},  and $\beta\propto1/T$. The density of states $\rho (E)$
is easily obtained from (\ref{PF}) by performing an inverse Laplace transform, so that one obtains
\begin{equation} \label{CompInt}
\rho(E) =\frac{1}{2 \pi i} \int^{a+ i\infty}_{a -i\infty}e^{S(\beta)}\text{d}\beta.
\end{equation}
Here, $S(\beta)$ is the entropy and its exact form is given in terms of the partition function and the total energy as $S(\beta)=\beta  E+\ln Z$, where $S_{0}=S(\beta_{0})$.
The  integral (\ref{CompInt}) can be evaluated using the method of  steepest decent,  around the saddle
point $\beta_0$, so that $[\partial S(\beta)/\partial\beta]_{\beta=\beta_0}$ vanishes, and the equilibrium relation
$E=-[\partial\ln Z(\beta)/\partial\beta]_{\beta=\beta_0}$ is satisfied. Therefore, the equilibrium temperature
is given by $T_{0}=1/\beta_0$, and we can expand the entropy $S(\beta)$
around the equilibrium point $\beta_0$ as follows
\begin{equation}\label{a1}
S(\beta)=S_0+\frac{1}{2}(\beta-\beta_0)^2 \left[\frac{\partial^2 S(\beta)}{\partial \beta^2 }\right]_{\beta=\beta_0}+\cdots.
\end{equation}
Here, the first term $S_0=S(\beta_0)$ denotes the entropy at the equilibrium, the second term represents the first order correction over
it. If we restrict ourselves to this first order and replace (\ref{a1}) into (\ref{CompInt}), we obtain
\begin{equation}
\rho(E) = \frac{e^{S_{0}}}{\sqrt{2\pi}} \left\{\left[\frac{\partial^2 S(\beta)}{\partial \beta^2 }\right]_{\beta = \beta_0}\right\}^{- \frac{1}{2}},
\end{equation}
for $[\partial^2 S(\beta)/\partial\beta^2]_{\beta=\beta_0}>0$, where we choose $a=\beta_0$ and $\beta-\beta_0=ix$ with $x$ being a real variable.
Thus, the expression of the microcanonical entropy $\mathcal{S}$ turns out to be \cite{1,2}
\begin{equation}\label{MicorEntropy}
\mathcal{S}=\ln\rho(E)=S_0-\frac{1}{2}\ln\left\{\left[\frac{\partial^2 S(\beta)}{\partial \beta^2 }\right]_{\beta = \beta_0}\right\}.
\end{equation}
Note that the entropy $S(\beta)$ given in (\ref{a1}) is different from $\mathcal{S}$ as given by (\ref{MicorEntropy}), the former $S(\beta)$ being
the entropy at any temperature, whereas the latter one, $\mathcal{S}$ is the corrected microcanonical entropy at equilibrium, which is computed by
incorporating small fluctuations around thermal equilibrium.
However, the result obtained in (\ref{MicorEntropy}) is completely model independent and, it
can be applied to any canonical thermodynamical system including a BIon solution. Thus, the first order correction is solely governed by the term
$\ln [\partial^2 S(\beta)/\partial\beta^2]_{\beta=\beta_0}$. This can be simplified to a generic form of
the entropy correction given by $\ln (CT^2)$ \cite{1, 2}.

Now for a system with equilibrium temperature  $\beta_0$, and the equilibrium entropy     $S_0$, the fluctuation around this equilibrium entropy,  $\ln [\partial^2 S(\beta) / \partial \beta^2 ]_{\beta = \beta_0}$ do  not depend on $\beta_0$ or $S_0$. However, the exact form of these  fluctuation can be obtained by assuming that this system is dual to a  conformal field theory, and using the    modular invariance of the  conformal field theory \cite{2, 3}. This is done by assuming that $S(\beta) = a \beta^m + b \beta^{-n},$ with  $m, n,  a/b >0$, and observing that at equilibrium this function has an extremum, with $\beta_0 = (nb/am)^{1/m+n}$. Using this observation, it can be demonstrated that this term $[\partial^2 S(\beta) / \partial \beta^2 ]_{\beta = \beta_0}$ (which represents first order correction around the equilibrium) has to be proportional to $\ln [S_0 \beta_0^2]$  \cite{1, 2, 3, 3a}. As the  Hawking temperature $T$ for the   black object is obtained at the  equilibrium, so we  identified  the equilibrium temperature    $\beta_0$  with $T$. Thus, the thermal fluctuations around the  equilibrium entropy can expressed in terms of the equilibrium entropy $S_0$ and equilibrium temperature $T$. It may be noted that such  corrections to the entropy of AdS black holes have been obtained  from the entropy of the  boundary theory using AdS/CFT correspondence \cite{x1, x2, x3, x4}, and it has been observed that the corrections to entropy can be expressed as a logarithmic function of the original equilibrium  temperature $T$.

As this correction  term is proportional to the logarithmic function of the area, it is of the  universal form of the leading order corrections to the entropy of the black object \cite{l1, l2, l3, l4, l5, l6}. It may be noted that even  thought the form of these corrections is universal, the exact value of the constant of  proportionality  to these corrections is model dependent
\cite{l1, l2, l3, l4, l5, l6}. Hence, as the constant of proportionality is model dependent, we will use a    free parameter $\gamma = [0,1]$ \cite{ y0, y1, y3, y4, y2}. Now it is obvious that, when we neglect these thermal  fluctuations $\gamma \to 0$, and we obtain the original equilibrium results. Furthermore, for $\gamma \neq 0 $, these corrections are proportional to the  logarithmic function of the area, and hence are the leading order corrections to  the entropy \cite{l1, l2, l3, l4, l5, l6}. Furthermore, as   these  are the leading order corrections to the equilibrium entropy, so they are expected to be  less the original equilibrium  entropy of the black object (for the perturbative expansion to be valid) \cite{ y0, y1,   y3, y4, y2}. It may be noted that when the black object is large, we can neglect thermal fluctuations, and at that stage, the original equilibrium  entropy can describe the system. Thus, $S_0$ is a good approximation to entropy when $S_0 >> ln (S_0 T^2)$. However, as the black object  reduces in size due to Hawking radiation, we need to consider the corrections to entropy due to thermal fluctuations. Thus, we need to consider  $\gamma \ln (S_0 T^2)$, when $ S_0 > \ln (S_0 T^2)$. These corrections are still smaller than the original equilibrium entropy,  but they are large enough to change the behavior of the system, and so they cannot be neglected. At first we can use the leading order corrections, and then we have to use higher order corrections. These corrections can be obtained by considering higher order perturbative corrections around the equilibrium \cite{m1,m2}. However, as the black hole approaches Planck scale and $S_0 \sim \ln (S_0 T^2)$,  the perturbative expansion  around equilibrium breakdown, and this expansion cannot be used to obtain the corrections to the entropy. This corresponds to the breaking of spacetime manifold by quantum fluctuations, which occur near Planck scale \cite{p1, p2}. So, this perturbative  approximation cannot be used for analyzing Planck scale black objects.

Now, as this expansion for leading order corrections,  holds for any black object whose degrees of freedom can be analyzed using a
CFT  \cite{2, 3, 3a}, and it has been argued that degrees of freedom of a BIon can also be analyzed  
using a CFT \cite{cft1y, cft2y}, we can use these corrections for analyzing a  BIon.
So,   we propose that the quantum correction to the entropy of a BIon can be expressed as
\begin{equation} \label{Corrected-Entropy}
\mathcal{S}=S_{0}-\frac{\gamma}{2}\ln{(S_0T^{2}) \mathcal{Y}} \sim S_{0}-\frac{\gamma}{2}\ln{(S_0T^{2}) - \frac{\gamma}{2} \mathcal{Y}},
\end{equation}
where $S_0$ is the original entropy of the BIon solution \cite{BIon01, bion}, and $\mathcal{Y}$  in general, is a function of other quantities (such as the properties of branes and string charges). Thus, a full analysis of this system should incorporate such a quantities, but as a toy model, we will only  analyze the corrections produced by $\gamma\ln{(S_0T^{2})}/2 $ on the thermodynamics of such a system, and neglect the effect of $\mathcal{Y}$. This can possible be justified by fixing certain quantities in the system, and analyzing it as a toy model. So, here the last term of (\ref{Corrected-Entropy}), along with the higher order corrections to entropy \cite{hnew} are neglected, and the leading order corrections to the entropy from thermal fluctuations  are considered.

It may be noted that such  logarithmic corrections terms  are universal, and occur in almost all approaches to quantum gravity. However,
the coefficient of such logarithmic correction  term is model dependent \cite{l1, l2, l3, l4, l5, l6}. As the corrected  expression used in this paper
involves a free parameter $\gamma =[0, 1]$  \cite{ y0, y1, y3, y4, y2}, it will  hold even using different approaches. As any  other approaches to this problem can   only change the value of this coefficient $\gamma$, which is not fixed in this paper. Thus, the validity of the (\ref{Corrected-Entropy})  can be argued on general grounds, and the main aim of the paper is to analyze the effects of such
a logarithmic corrections  on the thermodynamics of a BIon solution.

So, to obtain quantum corrections to the entropy of a BIon solution, we need to use the original entropy $S_0$ and original equilibrium temperature $T$ of the BIon solution \cite{BIon01, bion}.
Now a BIonic system is a flat spacetime configuration  of a D-brane  parallel to an anti-D-brane, connected by a wormhole, which has a F-string charge.
Geometrically, it is composed of $\mathcal{N}$ coincident D-branes which are infinitely extended, and has $\mathcal{K}$ units of F-string charge,
ending in a throat, with minimal radius $\sigma_0$  (at temperature $T$). To construct a wormhole solution from this, all we have to do, is to attach
a mirror solution at the end of the throat.

It is well known that the blackfold action can be used to describe the D-brane for probing the zero temperature background.
However, it was shown that one can also use DBI action for probing the thermal backgrounds \cite{bion}, where it is ensured that the brane is not affected by
the thermal background, but the degrees of freedom living on the brane are 'warmed up' due to the temperature of thermal background. Thus, the thermal background
acts as a heat bath to the D-brane probe, and due to this, the probe stays in thermal equilibrium with the thermal background, which is a ten dimensional hot flat space.
This is constructed in the blackfold approach. Thus the thermal generalization of BIon solution has  been carried out, and the thermodynamic quantities for this
configuration are given by \cite{BIon01, bion}
\begin{eqnarray}\label{F}
M &=& \frac{4T^{2}_{D3}}{\pi T^{4}}\int_{\sigma_{0}}^{\infty}d\sigma \frac{\sigma^{2}(4cosh^{2}\alpha +1)F(\sigma)}{\sqrt{F^{2}(\sigma)-F^{2}(\sigma_{0})}\cosh^{4}\alpha}, \\
S_{0} &=& \frac{4T^{2}_{D3}}{\pi T^{5}}\int_{\sigma_{0}}^{\infty}d\sigma \frac{4\sigma^{2}F(\sigma)}{\sqrt{F^{2}(\sigma)-F^{2}(\sigma_{0})}\cosh^{4}\alpha },  \label{a} \\
\mathcal{F} &=& \frac{4T^{2}_{D3}}{\pi T^{4}}\int_{\sigma_{0}}^{\infty}d\sigma \sqrt{1+z'^{2}(\sigma)} F(\sigma),
\end{eqnarray}
where $M$ is the total mass, $S_0$ is the  entropy and $\mathcal{F}$ is  free energy of the BIon. Here, $T_{D3}$ is the D3-brane tension, $z$ is a 
transverse coordinate to the branes and $F(\sigma)=\sigma^{2}(4\cosh^{2}\alpha-3)/\cosh^{4}\alpha$, with $\sigma$ being the world
volume coordinate and $\sigma_{0}$ being the minimal (sphere) radius of the throat or wormhole. Here $\alpha$ is a function of the temperature.
The chemical potentials for the D3-brane and F-string are as follows
\begin{eqnarray}
\mu_{D3} &=& 8\pi T_{D3}\int_{\sigma_{0}}^{\infty}d\sigma \frac{\sigma^{2} \tanh \alpha \cos \zeta F(\sigma)}{\sqrt{F^{2}(\sigma)-F^{2}(\sigma_{0})}}  \\
\mu_{F1} &=& 2 T_{F1}\int_{\sigma_{0}}^{\infty}d\sigma \frac{\tanh \alpha \cos \zeta F(\sigma)}{\sqrt{F^{2}(\sigma)-F^{2}(\sigma_{0})}}
\end{eqnarray}	
It should be noted that these relations satisfy the first law of thermodynamics $dM=TdS_{0}+\mu_{D3}d\mathcal{N} +\mu_{F1}d\mathcal{K}$
as well as the Smarr relation, $4(M-\mu_{D3} \mathcal{N}-\mu_{F1} \mathcal{K})-5TS_{0}=0$. One can also calculate internal energy and the
specific heat of the BIon solution as
\begin{alignat}{1}
U_{0} &=\frac{4T^{2}_{D3}}{\pi T^{4}}\int_{\sigma_{0}}^{\infty}d\sigma  F(\sigma)\left[\sqrt{1+z'^{2}(\sigma)}+\frac{4\sigma^{2}}{\sqrt{F^{2}(\sigma)-F^{2}(\sigma_{0})}\cosh^{4}\alpha}\right], \\
C_{0} &= T\bigg(\frac{dS_{0}}{dT}\bigg)=-\frac{20T^{2}_{D3}}{\pi T^{5}}\int_{\sigma_{0}}^{\infty}d\sigma \frac{4\sigma^{2}F(\sigma)}{\sqrt{F^{2}(\sigma)-F^{2}(\sigma_{0})}\cosh^{4}\alpha},
\end{alignat}
which indicates that the system has a negative specific heat.
\section{Corrected Thermodynamics for the BIon}\label{sec4}
 Let us now look for the thermal corrections to the above equations by considering logarithmic correction to the
entropy $S$ given by the equation (\ref{Corrected-Entropy}). The entropy (\ref{a}) of $\mathcal{N}$ coincident D-branes, with a throat solution gets corrected as
\begin{eqnarray}\label{Corrected-Entropy-222}
S&=&\frac{4T^{2}_{D3}}{\pi T^{5}}\int_{\sigma_{0}}^{\infty}d\sigma \frac{4\sigma^{2}F(\sigma)}{\cosh^{4}\alpha\sqrt{F^{2}(\sigma)-F^{2}(\sigma_{0})}}\nonumber\\
&-&\frac{\gamma}{2} \ln \left[\frac{4T^{2}_{D3}}{\pi T^{3}}\int_{\sigma_{0}}^{\infty}d\sigma \frac{4\sigma^{2}F(\sigma)}{\cosh^{4}\alpha\sqrt{F^{2}(\sigma)-F^{2}(\sigma_{0})}}\right].
\end{eqnarray}
In order to analyze the expression for the corrected entropy, we can assume $\cosh^{2}\alpha(\sigma_{0})=\frac{3}{4}$,
which means that $F(\sigma_{0})=0$ produces a relatively easy solution. It is indeed a possible solution of the equation of motion at $\sigma_{0}$ \cite{BIon01}. However, we would like to work on the regime where
the branch is connected to the extremal BIon \cite{BIon01}. In this formulation one obtains
\begin{figure}
 \begin{center}$
 \begin{array}{cccc}
\includegraphics[width=54 mm]{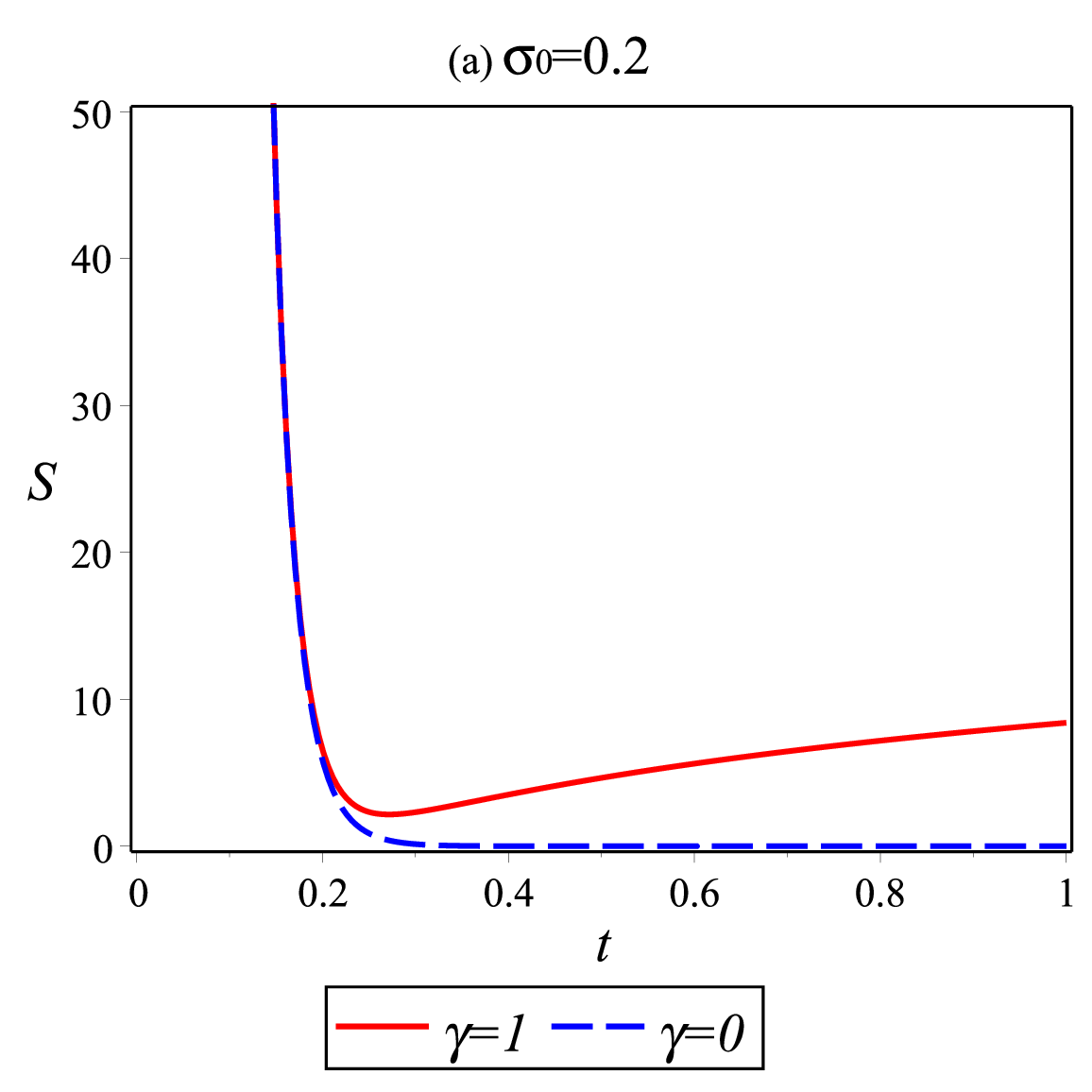}\includegraphics[width=54 mm]{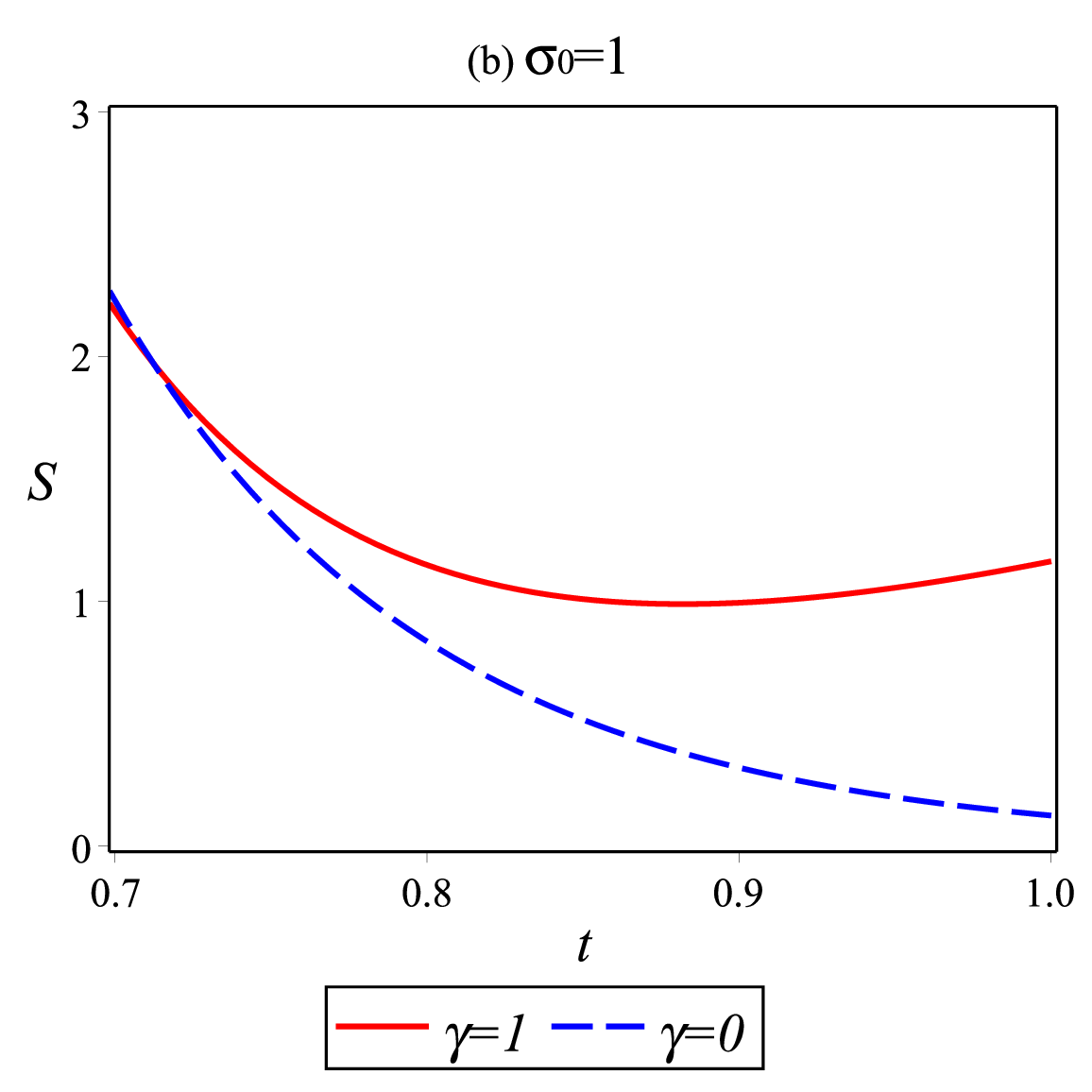}\includegraphics[width=54 mm]{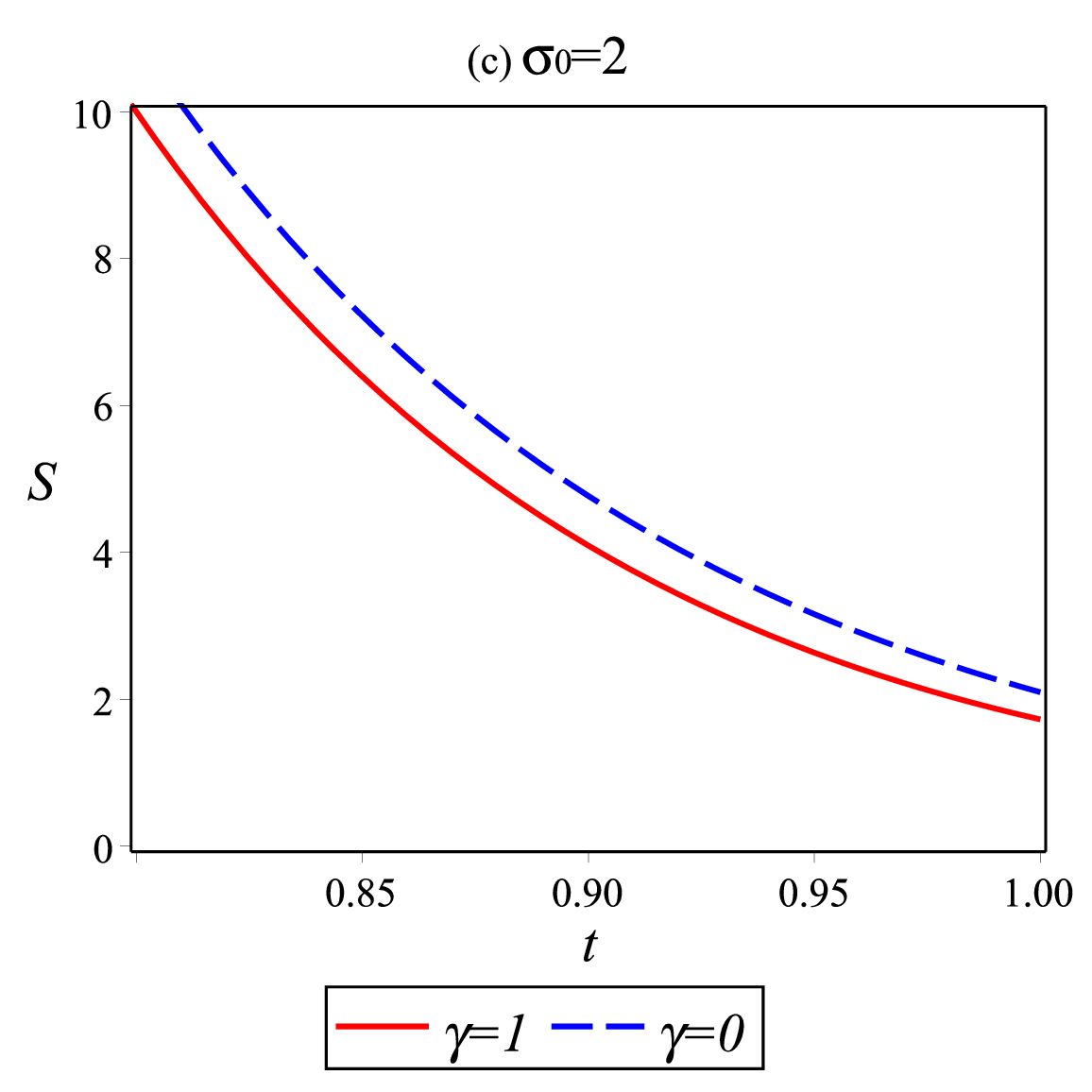}
 \end{array}$
 \end{center}
\caption{\small{Behavior of the corrected entropy as a function of $t\equiv\bar{T}$ for $\mathcal{K}=1$ and $T_{D3}=1$.}}
 \label{fig1}
\end{figure}

\begin{figure}[!b]
 \begin{center}$
 \begin{array}{cccc}
\includegraphics[width=75 mm]{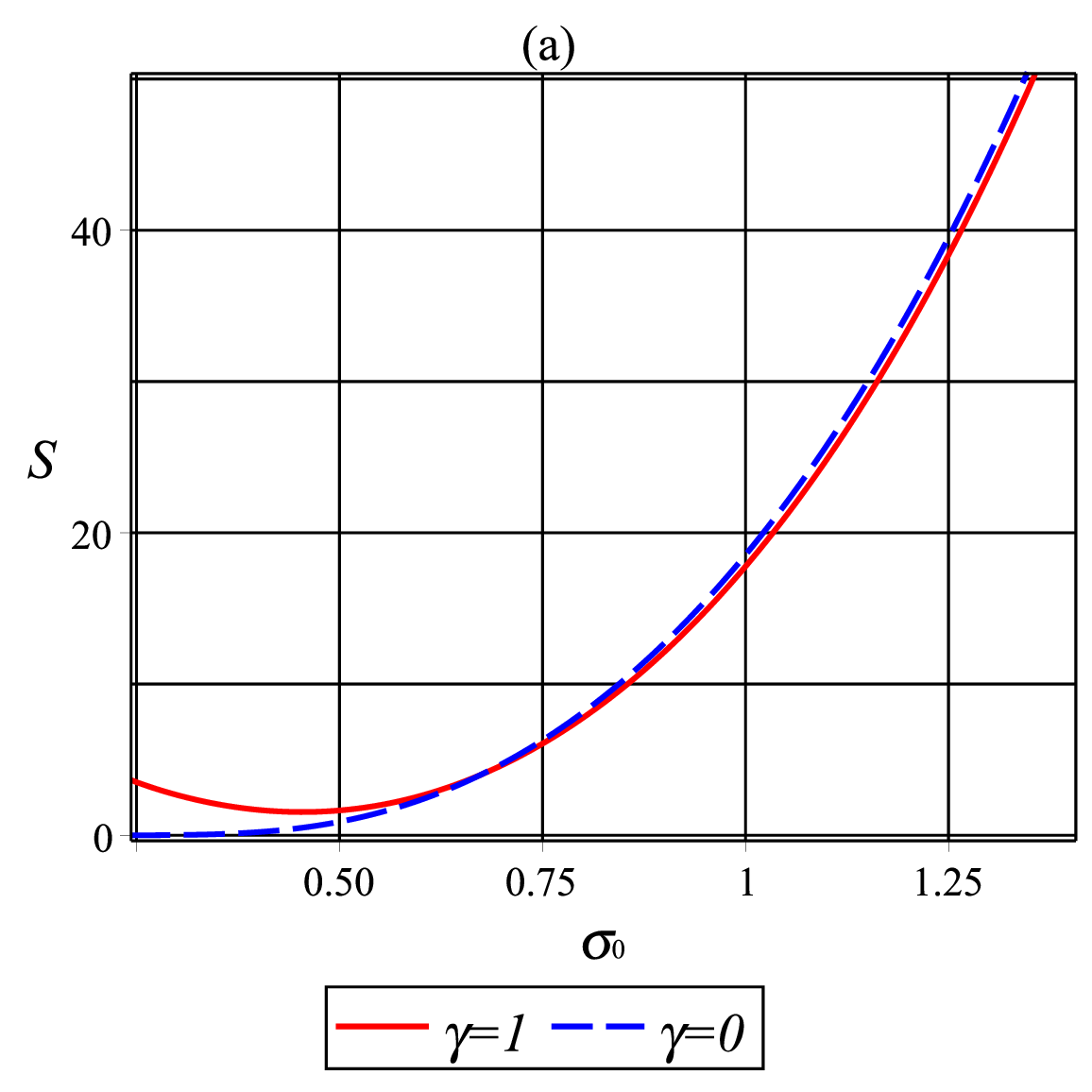}\includegraphics[width=75 mm]{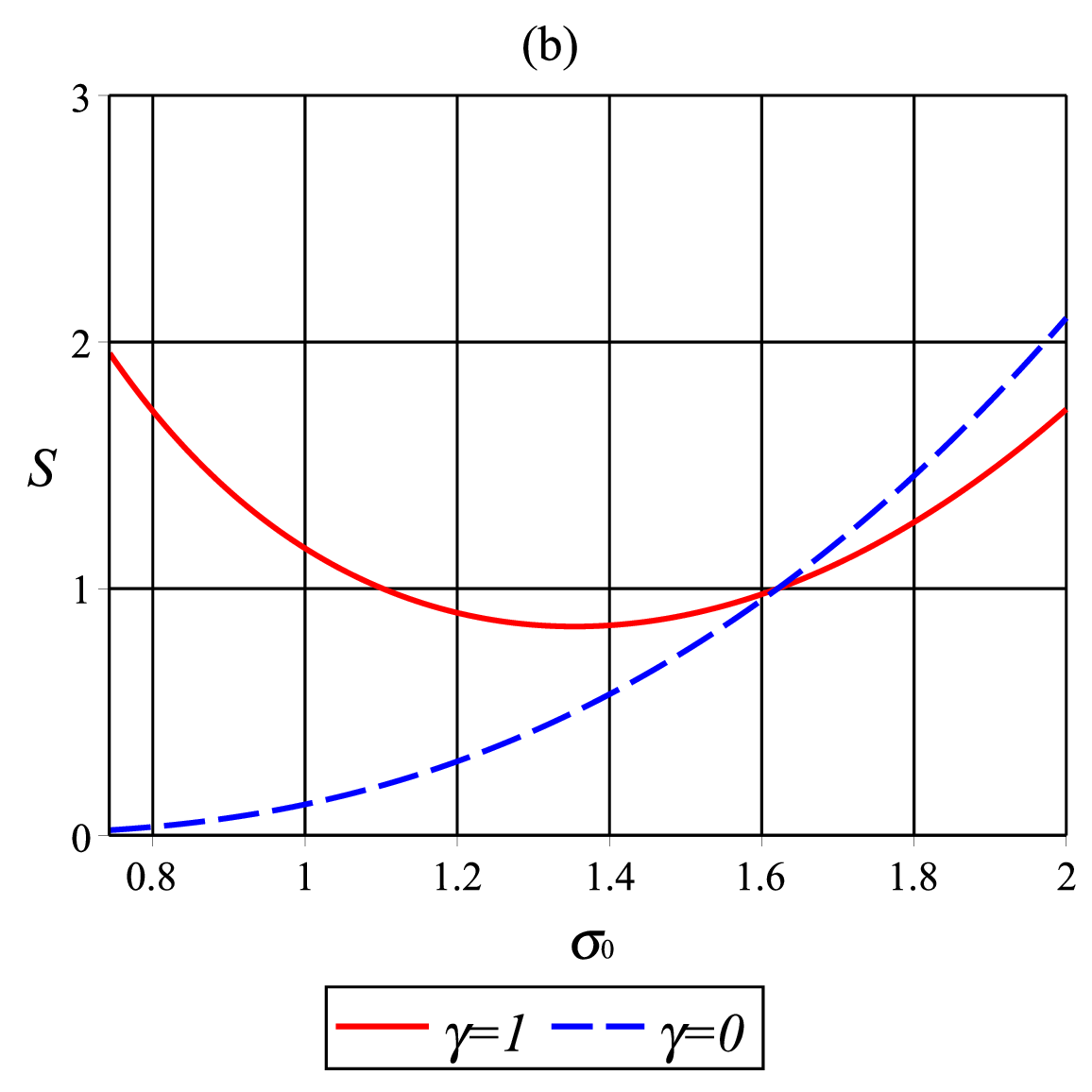}
 \end{array}$
 \end{center}
\caption{\small{Behavior of the corrected entropy as a function of $\sigma_{0}$ for $\mathcal{K}=1$, $T_{D3}=1$ (a) $\bar{T}=0.5$ (b) $\bar{T}=1$.}}
 \label{fig2}
\end{figure}
\begin{figure}
 \begin{center}$
 \begin{array}{cccc}
\includegraphics[scale=0.4]{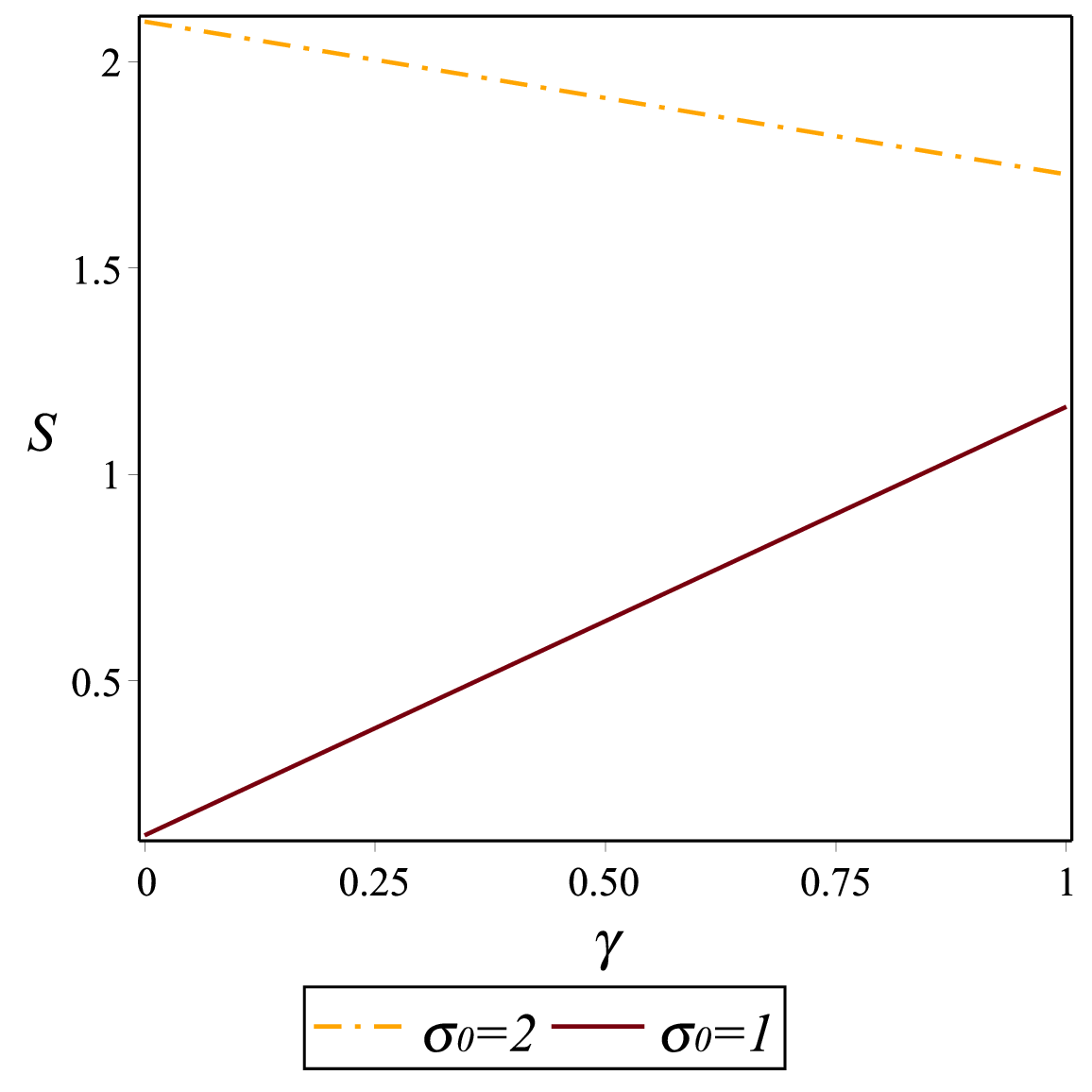}
 \end{array}$
 \end{center}
\caption{\small{Behavior of the corrected entropy in terms of $\gamma$ to see cases of $\gamma=0$ and $\gamma=1$. We set $\mathcal{K}=1$, $\bar{T}=1$ and $T_{D3}=1$.}}
 \label{fig3}
\end{figure}

\begin{equation}
\cosh^{2}\alpha=\frac{3}{2}\frac{\cos{\frac{\delta}{3}}+\sqrt{3}\cos{\frac{\delta}{3}}}{\cos\delta},
\end{equation}
where $\cos\delta={\bar{T}}^{4}\sqrt{1+\mathcal{K}^2/\sigma^{4}}$, with ${\bar{T}}^{4}=9\pi^{2}T^4\mathcal{N}/(4\sqrt{3}T_{D3})$.
We further assume an infinitesimal $\delta$ (corresponding to $\sigma^{2}>\mathcal{K}$ at low temperature $\bar{T}\approx1$),
so that $\cos\delta\approx1$ and $\sin\delta\approx\delta\approx{\bar{T}}^{4}[1+1/(2\sigma^{4})]$. Here we have
considered $\mathcal{K}=1$, as it can always be absorbed in $\sigma$ by means of a re-scaling. Now the remaining parameter is $T_{D3}$, which we can set to one,  
$T_{D3}=1$. So, we can  plot the corrected entropy (\ref{Corrected-Entropy-222}) by varying the $\bar{T}$, $\sigma_{0}$ and $\gamma$ as depicted in
Figs. \ref{fig1}, \ref{fig2} and \ref{fig3}. In Fig. \ref{fig1}, we draw the corrected entropy in terms of $\bar{T}$ for different values
of $\sigma_{0}$. We should note that there is a minimal radius $\sigma_{min}={\bar{T}}^{2}(1-{\bar{T}}^{8})^{-\frac{1}{4}}$ corresponding to
each plots, with $\sigma_{0}\geq \sigma_{min}$. In the panel (a) of Fig. \ref{fig1}, we have considered $\sigma_{0}=0.2$, such that ${\bar{T}}\geq0.45$ and,
in such a situation, we can see that the corrected entropy is larger than the uncorrected one. This means that the logarithmic corrections have increased the value
of the entropy. Similar thing happens in the panel (b), where $\sigma_{0}=1$, thus, ${\bar{T}}\geq0.9$. However, the system behaves   differently,  when
we consider $\sigma_{0}=2$ and ${\bar{T}}\geq0.99$, as shown in Fig. \ref{fig1}(c). Here we see that corrected entropy is smaller than the uncorrected entropy.
Therefore, the behavior of the entropy with the logarithmic correction depends on both the parameters, namely, the temperature and $\sigma_{0}$.

In Fig. \ref{fig2}, we plot the corrected entropy as a function of $\sigma_{0}$ for ${\bar{T}}\geq0.5$, i.e. $\sigma_{min}=0.25$. In this case,
we see that there exists a critical $\sigma_{c}$ for each of the plots, and when the value of $\sigma_{0}$ is less than the value of $\sigma_{c}$,
the corrected entropy is larger than the uncorrected one. Whereas, when $\sigma_{0}>\sigma_{c}$, the corrected entropy is less. However, the critical
points depend on the temperature, for instance, in the case when ${\bar{T}}=0.5$, i.e. in Fig. \ref{fig2}(a), we notice that $\sigma_{c}\approx0.7$,
while in Fig. \ref{fig2}(b), i.e. for ${\bar{T}}\approx1$, $\sigma_{c}\approx1.6$. We should note that the region compatible with our assumption is
$\sigma_{0}>1$. Fig.\,\ref{fig3} demonstrates the behavior of the entropy with respect to $\gamma$, which is the coefficient that determines the amount
of corrections on the system. By choosing ${\bar{T}}\geq1$, we can see that the entropy is a decreasing function of $\gamma$ for $\sigma_{0}=2$,
while it is an increasing function of $\gamma$ for $\sigma_{0}=1$. It means that for the small throat (smaller than $\sigma_{c}$), the thermal fluctuation
 increase the entropy, which may yield more stability to the system, with maximum value of the entropy. On the other hand, a bigger throat may make the system unstable.

The logarithmic correction also modifies the internal energy and the specific heat. Let us analyze the effects for the modification
of the internal energy first, which we compute as follows
\begin{figure}[!b]
 \begin{center}$
 \begin{array}{cccc}
\includegraphics[width=75 mm]{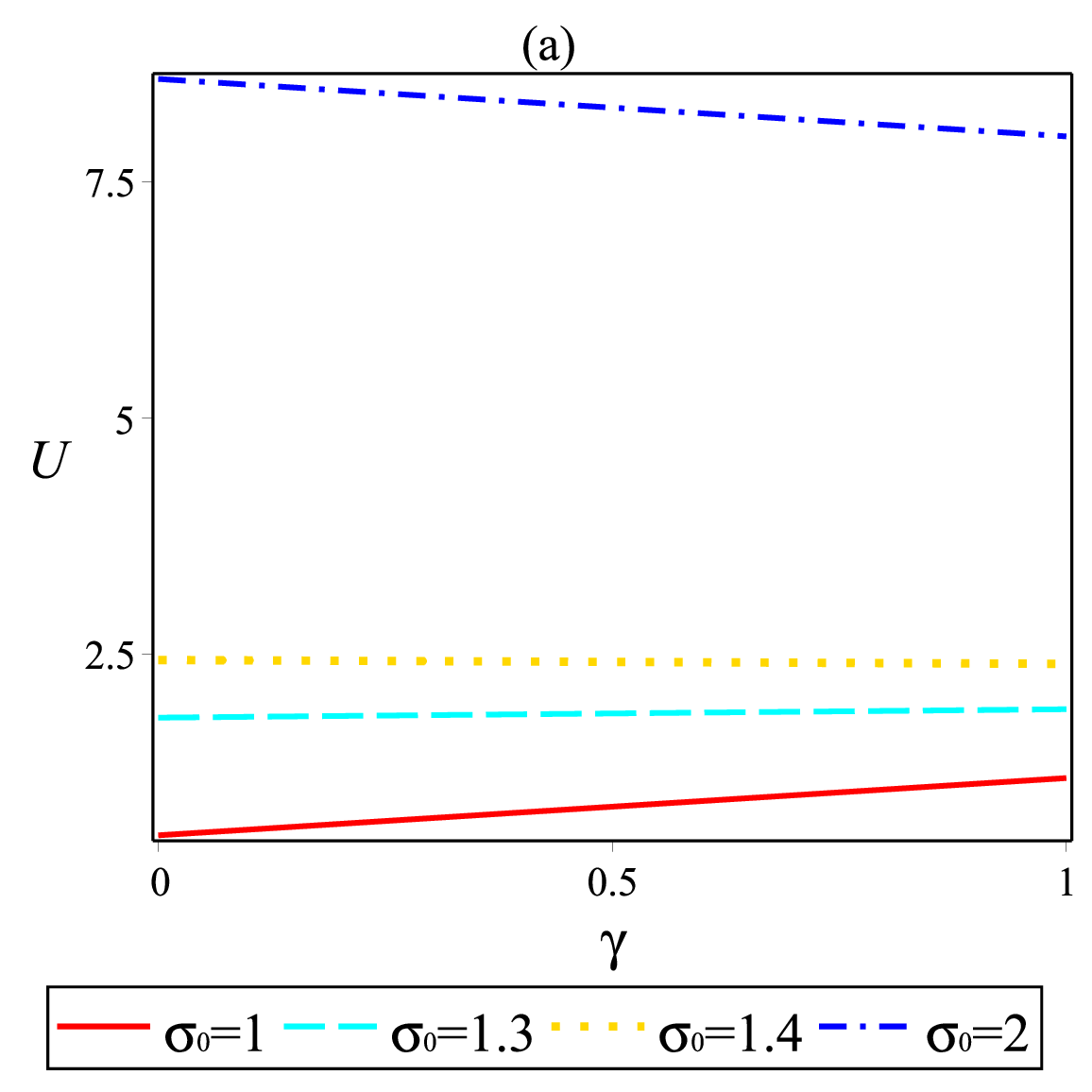}\includegraphics[width=75 mm]{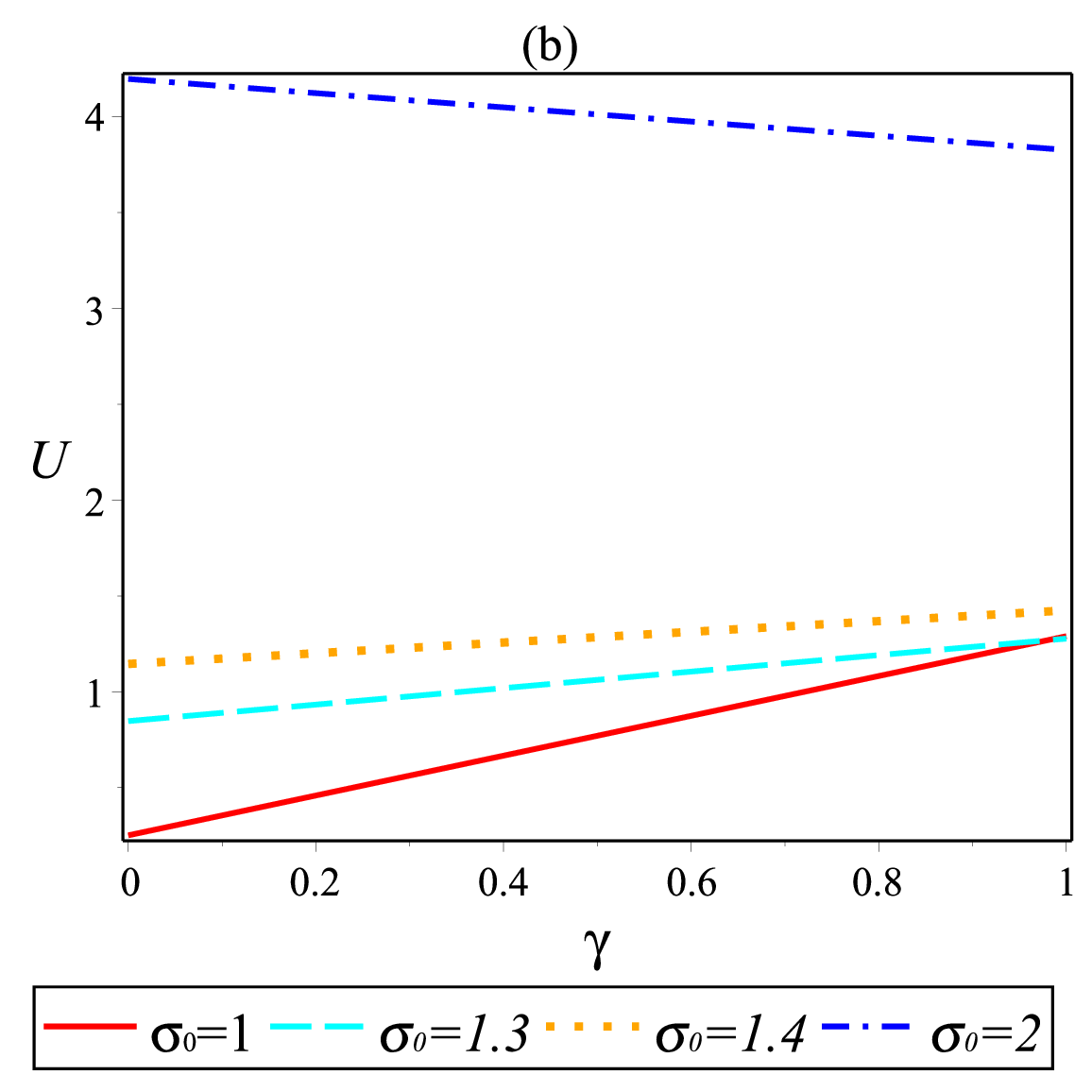}
 \end{array}$
 \end{center}
\caption{\small{Behavior of internal energy in terms of $\gamma$, with $\mathcal{K}=1$ and $T_{D3}=1$ (a) $\bar{T}=0.9$ (b) $\bar{T}=1$.}}
 \label{fig4}
\end{figure}

\begin{figure}
 \begin{center}$
 \begin{array}{cccc}
\includegraphics[width=75 mm]{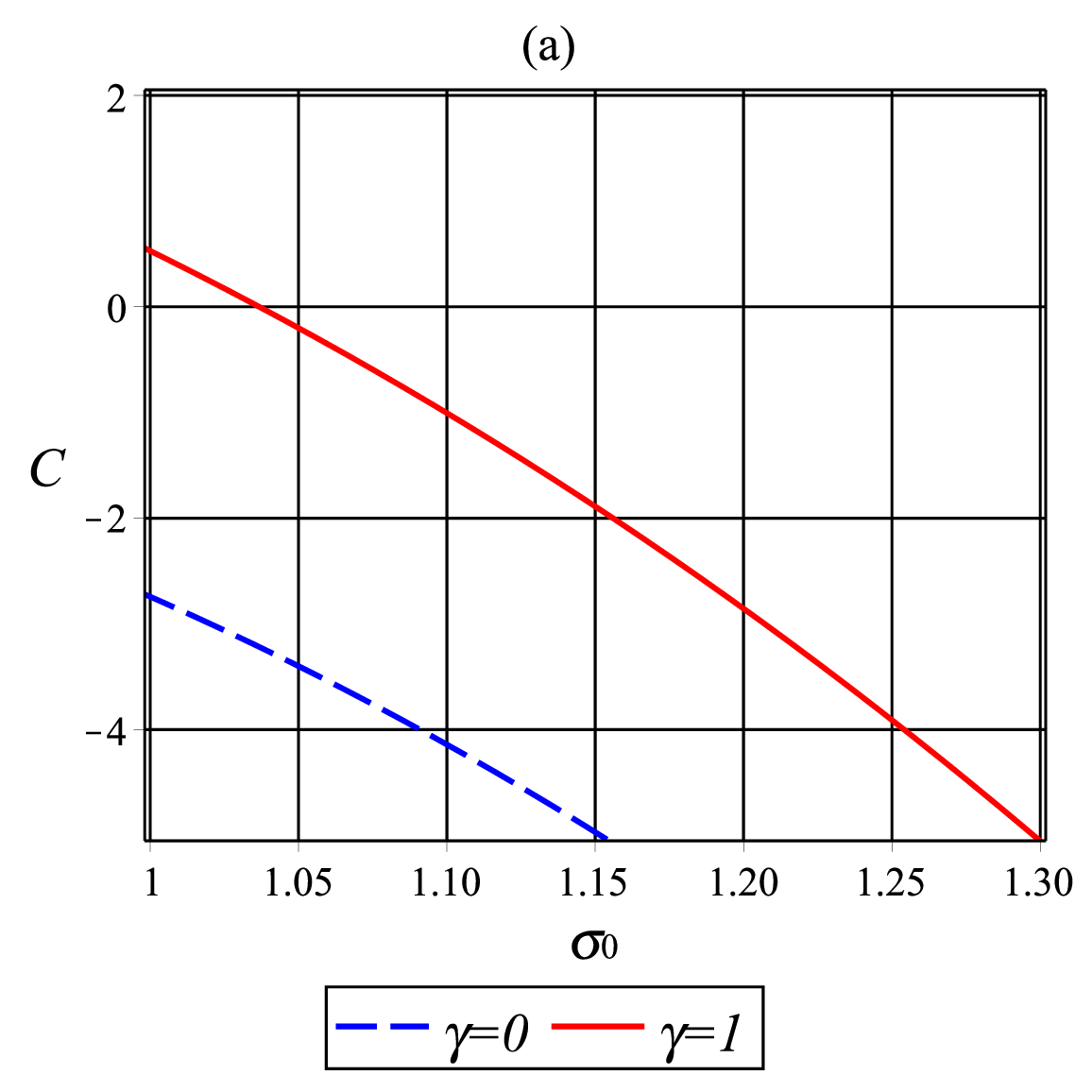}\includegraphics[width=75 mm]{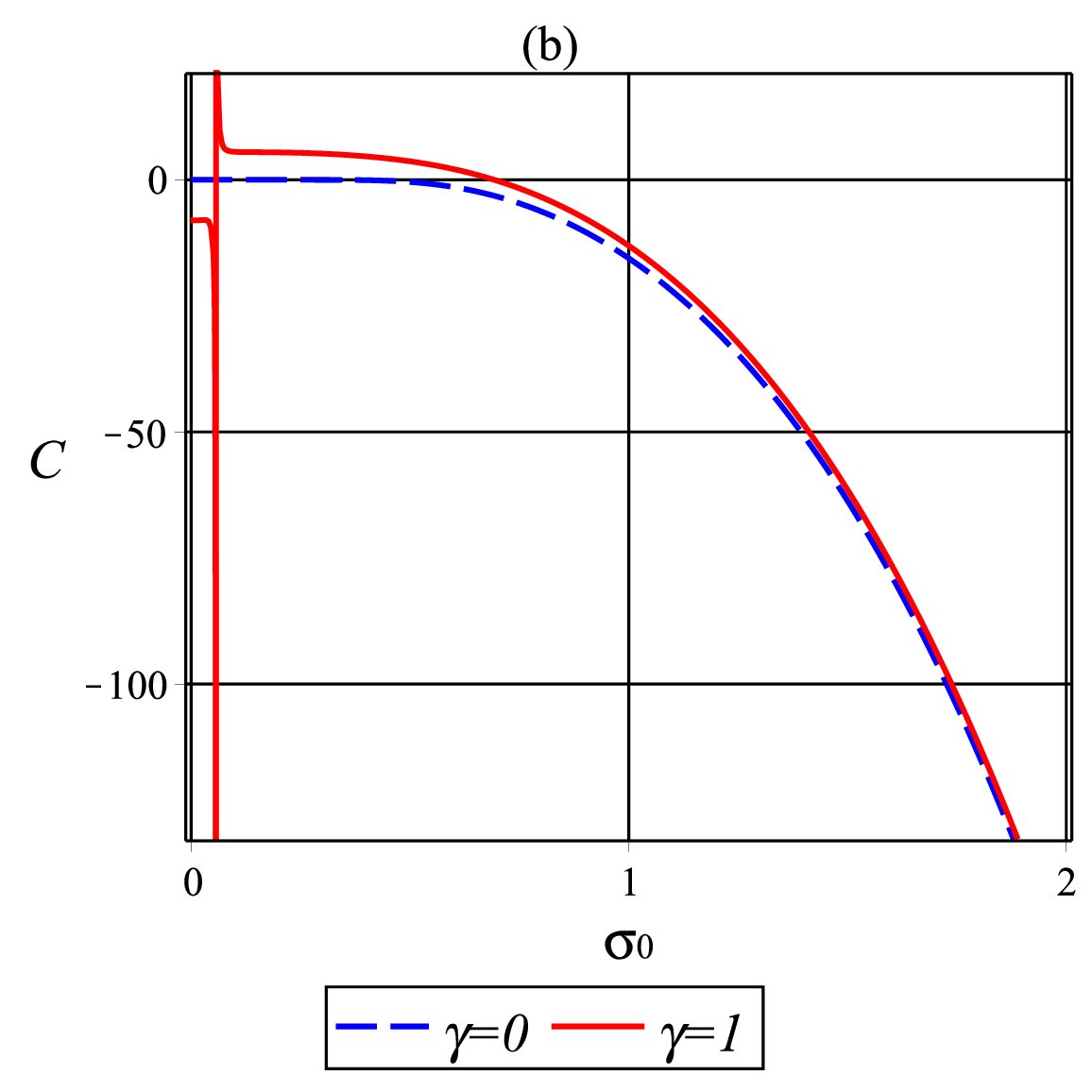}
 \end{array}$
 \end{center}
\caption{\small{Behavior of the specific heat with respect $\sigma_{0}$ for $\mathcal{K}=1$ and $T_{D3}=1$ (a) $\bar{T}=0.9$ (b) $\bar{T}=0.7$.}}
 \label{fig5}
\end{figure}

\begin{figure}[!b]
 \begin{center}$
 \begin{array}{cccc}
\includegraphics[scale=0.4]{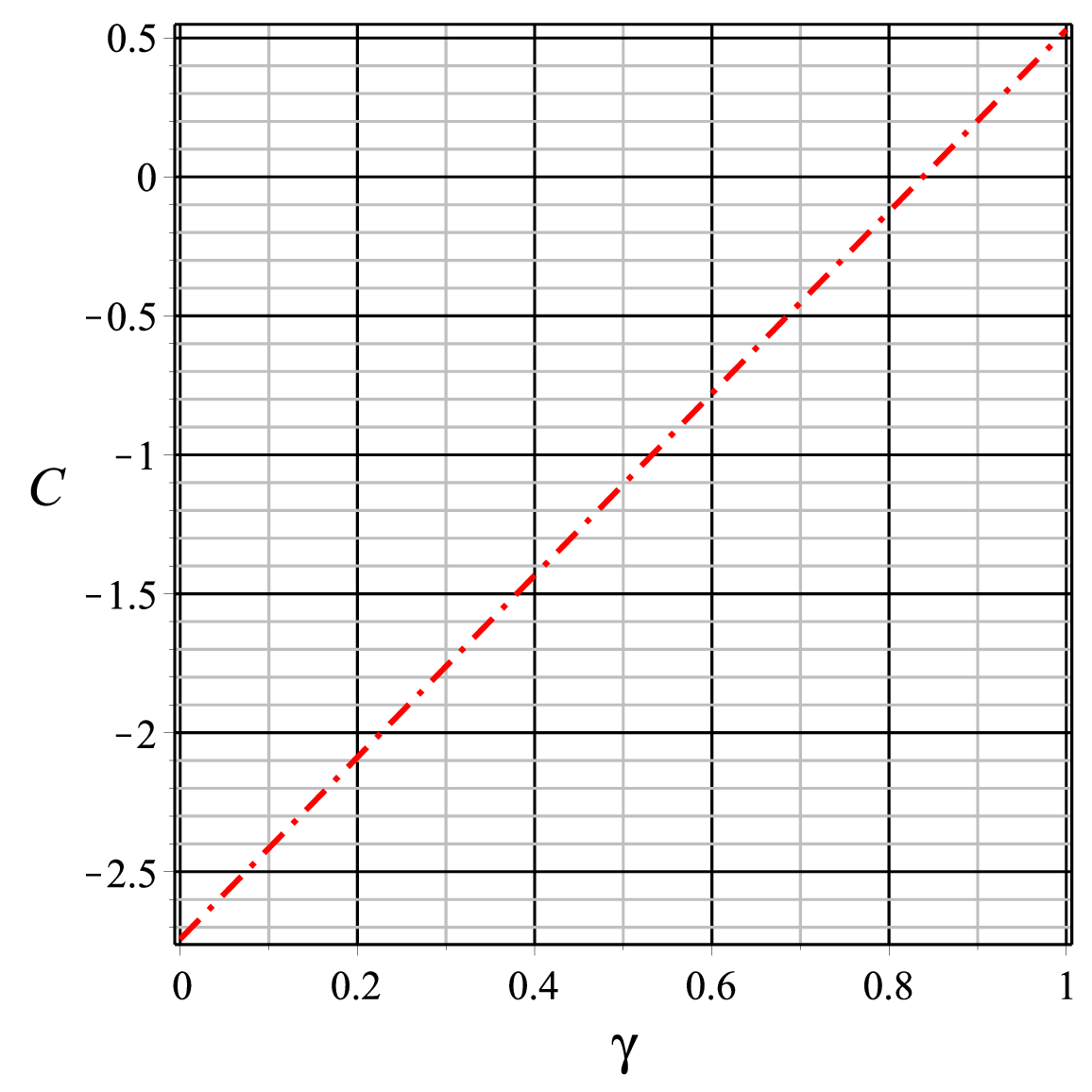}
 \end{array}$
 \end{center}
\caption{\small{Behavior of the specific heat with respect to $\gamma$ for $\mathcal{K}=1$, $\bar{T}=0.9$ and $T_{D3}=1$.}}
 \label{fig6}
\end{figure}

\begin{figure}[!b]
 \begin{center}$
 \begin{array}{cccc}
\includegraphics[scale=0.5]{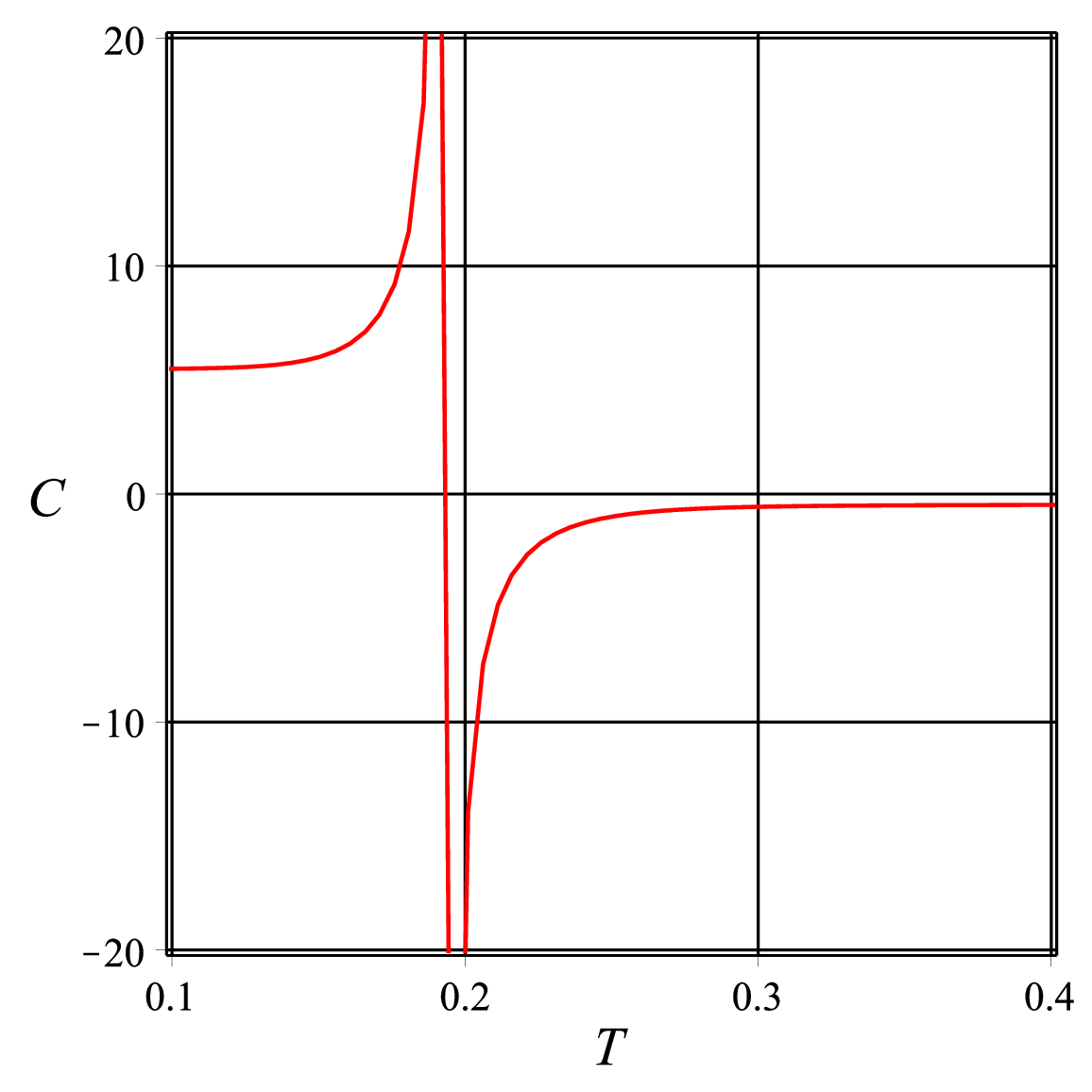}
 \end{array}$
 \end{center}
\caption{\small{Behavior of the specific heat with respect to $\bar{T}$ for $\mathcal{K}=1$, $\sigma_{0}=0.02$ and $T_{D3}=1$.}}
 \label{fig7}
\end{figure}

\begin{align}
U&=\frac{4T^{2}_{D3}}{\pi T^{4}}\int_{\sigma_{0}}^{\infty}d\sigma F(\sigma)\left[\sqrt{1+z'^{2}(\sigma)}+\frac{4\sigma^{2}}{\sqrt{F^{2}(\sigma)-F^{2}(\sigma_{0})}\cosh^{4}\alpha}\right] \nonumber \\
&~~~~-\frac{\gamma T}{2} \ln \left[\frac{4T^{2}_{D3}}{\pi T^{3}}\int_{\sigma_{0}}^{\infty}d\sigma \frac{4\sigma^{2}F(\sigma)}{\sqrt{F^{2}(\sigma)-F^{2}(\sigma_{0})}\cosh^{4}\alpha}\right].
\end{align}
We can perform a graphical analysis similar to the entropy to give similar results. However, we focus only on the behavior of the
internal energy with the variation of the parameter $\gamma$. Fig. \ref{fig4} shows that, although the variation of the internal
energy is smaller with the logarithmic corrections, however, the slope of the increasing or decreasing functions depends on the value
of the temperature and the radius of the throat, as expected. For larger radius, the entropy is decreased due to the thermal fluctuations,
while for the smaller radius the entropy is increased. Let us now see the effects on the specific heat. The exact expression of the corrected specific heat is given by
\begin{align}\label{heat}
C=T\frac{d}{dT}\left(S-\frac{\gamma}{2}\ln [ST^{2}]\right),
\end{align}
which has been analyzed in Fig. \ref{fig5}. Here, we show the effect of the logarithmic correction on the specific heat inside the
allowed region $1\leq\sigma_{0}$ for Fig. \ref{fig5}(a), and for the whole range in Fig. \ref{fig5}(b) in order to explore the general
approximate behavior. In the case of $\gamma=0$, we find that the specific heat is entirely negative, however, in the presence of the thermal
fluctuations there are some regions where it is positive. This means that in the presence of the logarithmic correction, there is a special radius $\sigma_{s}$
for which the specific heat is negative, $\sigma_{0}>\sigma_{s}$, while it is positive for $\sigma_{0}<\sigma_{s}$. Here again  the value of the $\sigma_{s}$
depends on the temperature, for instance, in Fig. \ref{fig5}(a) it is obvious that when we choose ${\bar{T}}=0.9$, we obtain $\sigma_{s}\approx1.03$. On the
other hand, in Fig. \ref{fig5}(b), when we increase $\sigma_{0}$, the difference between the corrected and uncorrected case slowly vanishes. It indicates that
the thermal fluctuation becomes relevant for the smaller radius. Also, we can see from the Fig. \ref{fig5}(b) an asymptotic behavior which may be interpreted as a first order phase transition as found in \cite{BIon01}. Here we show that it occurs due to the thermal fluctuations. We can confirm this by analyzing the free energy ($\mathcal{F}$) and find that $\frac{\partial\mathcal{F}}{\partial {\bar{T}}}=0$ at the phase transition point. It is indicated that we have a first order phase transition in this system. 

Also, we demonstrate the variation of the specific heat
with the parameter $\gamma$ in Fig. \ref{fig6}. It is obvious that the effect of the logarithmic correction is to increase the specific heat and, for $\gamma>0.65$
(approximately) the specific heat is completely positive, while for $\gamma=0$ it is completely negative. Finally in the 
Fig. \ref{fig7}, we can analyze the variation of the specific heat with ${\bar{T}}$ to plot its asymptotic behavior. 
At this asymptotic point, the first derivative of the free energy with respect to the temperature is zero, and so it is a first order phase transition.
\section{Conclusion}
Quantum fluctuations are  important when  dealing with objects of  very small length scales. 
They can be neglected when the object is large   compared to the Planck scale, but for small objects, the quantum fluctuation become important. Now at Planck scale the  background spacetime breaks down, and it is difficult to analyze this system. However, there is a stage before such a total breakdown, when the quantum fluctuations can be not be neglected, but can be analyzed as perturbations around a fixed spacetime. This corresponds to analyzing thermal fluctuations around 
equilibrium  for a black hole.  
We analyze the quantum corrections for the BIonic systems using such fluctuations around  equilibrium. We explicitly include the correction terms produced by such thermal fluctuations. We also analyze the behavior of system at  
the critical points, after  including these correction terms. Moreover, we demonstrate that these correction terms are important  by analyzing the stability conditions for this system.  In fact, we can demonstrate numerically that  these quantum fluctuation  affects the critical points, thus affecting  the stability of the system. The stability  increases under certain  conditions.  For instance, when the throat is smaller, the inclusion of
fluctuations increases the stability. Our analysis explicitly shows how the quantum fluctuation terms become important, with the decrease of the radius.
In  Fig. \ref{fig5}(b), we observe  that  increasing the radius $\sigma_0$, the correction term slowly vanishes, and the corrected result become identical with the uncorrected one. Apart from the stability analysis, we have computed the corrections to the internal energy and specific heat due to the quantum fluctuation. We confirmed that the  first order phase transition occurs in this system, by analyzing the free energy.
We have also analyzed the change in the behavior of the corrected system with the temperature. 

It may be noted that here we have only analyzed the system numerically. However, to demonstrate that the system actually have phase transition, it is important to analyze it analytically. It would thus be  important to analyze this system analytically. However, as this system is very complicated, it might be interesting to study a simpler model of black branes analytically. Such an analysis can give a better understating of such a system. 
It is possible to construct a BIon solution in M-theory using a system of M2-branes and M5-branes \cite{m22}.
It would be interesting to analyze the system at a finite temperature. Then the thermodynamics of this system
can be studied. It would be possible to study the quantum fluctuations to the geometry of a BIon in M-theory,
and these could produce thermal fluctuations in the thermodynamics of this system. It would be interesting
to analyze the critical points for such a system, and study the effects of these fluctuations on the
stability of this system. It may also be noted that the thermodynamics of  the  AdS black hole
have been studied in M-theory \cite{iib1, iib2}. It is possible to analyze the quantum corrections to these black holes,
and this can also be done in Euclidean Quantum Gravity. It would also be interesting to generalize the work
of this paper to such AdS black holes in M-theory.\\\\

\noindent \textbf{\large{Acknowledgments:}} S.D. acknowledges the support of research grant (DST/INSPIRE/04 /2016/001391) from the Department
of Science and Technology, Govt. of India. S.C. is funded by Conicyt grant 21181211. The Centro de Estudios Cient\'{\i}ficos (CECs) is funded by the Chilean Government through the Centers of Excellence Base Financing Program of Conicyt.


\end{document}